\documentclass{emulateapj}
\def\stacksymbols #1#2#3#4{\def\theguybelow{#2}
        \def\verticalposition{\lower#3pt}
        \def\spacingwithinsymbol{\baselineskip0pt\lineskip#4pt}
        \mathrel{\mathpalette\intermediary#1}}
\def\intermediary #1#2{\verticalposition\vbox{\spacingwithinsymbol
        \everycr={}\tabskip0pt
        \halign{$\mathsurround0pt#1\hfil##\hfil$\crcr#2\crcr
                \theguybelow\crcr}}}
\def\lta{\stacksymbols{<}{\sim}{2.5}{.2}}

\shorttitle{COSMIC RAY DIFFUSION}
\shortauthors{MATHEWS \& GUO}

\begin{document}

\title{COSMIC RAY DIFFUSION FRONTS IN THE VIRGO CLUSTER}

\author{William G. Mathews\altaffilmark{1} and
Fulai Guo\altaffilmark{1}}

\altaffiltext{1}{University of California Observatories/Lick
Observatory,
Department of Astronomy and Astrophysics,
University of California, Santa Cruz, CA 95064
mathews@ucolick.org}

\begin{abstract}
The pair of large radio lobes in the Virgo cluster, each about 23 kpc
in radius, have curiously sharp outer edges where the
radio-synchrotron continuum flux declines abruptly.  However, just
adjacent to this sharp transition, the radio flux increases.  This
radio limb-brightening is observed over at least half of the perimeter
of both lobes.  We describe slowly propagating steady state diffusion
fronts that explain these counterintuitive features.  Because of the
natural buoyancy of radio lobes, the magnetic field is largely tangent
to the lobe boundary, an alignment that polarizes the radio emission
and dramatically reduces the diffusion coefficient of relativistic
electrons.  As cosmic ray electrons diffuse slowly into the cluster
gas, the local magnetic field and gas density are reduced as 
gas flows back toward the radio lobe. Radio emission peaks can occur
because the synchrotron emissivity increases with magnetic field and
then decreases with the density of non-thermal electrons.  A detailed
comparison of steady diffusion fronts with quantitative radio
observations may reveal information about the spatial variation of
magnetic fields and the diffusion coefficient of relativistic
electrons.  On larger scales, some reduction of the gas density inside
the Virgo lobes due to cosmic ray pressure must occur and may be
measurable.  Such X-ray observations could reveal important information
about the presence of otherwise unobservable non-thermal components
such as relativistic electrons of low energy or proton cosmic rays.
\end{abstract}

\vskip.1in

\section{Introduction}
Several nearby galaxy clusters contain extended non-thermal radio lobes that
appear to be unrelated to currently visible X-ray cavities.
For example, the iconic image of the Virgo cluster 
at 90 cm (325 MHz) from Owen, Eilek \& Kassim (2000), shown in Figure 1, 
features two large opposing quasi-circular radio lobes.
(The famous 2 kpc radio jet in M87 is over-exposed in the central
dark region.)
In Mathews \& Brighenti (2008a) we proposed that the Virgo
radio lobes were formed by relativistic electrons that diffused through the
walls of X-ray cavities that have long-since buoyantly risen
and disappeared from view. 
This conjecture seemed attractive because 
a 25-kpc long radial thermal X-ray filament 
(Young, Wilson \& Mundell 2002; Forman, et al. 2005;
Werner, N. et al. 2010; Million et al. 2010) passes in projection 
right through the center of the southern Virgo lobe, 
suggesting that both were formed in the same X-ray cavity event.
Moreover, the radio-synchrotron
spectral age of the oldest and most distant electrons
in the Virgo lobes, $t_{sync} \approx 10^8$ yrs,  matches
the dynamical age of the post-cavity filament 
as computed by Mathews \& Brighenti (2008a) who 
suggested that the radio-synchrotron electrons currently 
filling the southern lobe 
may have diffused through the walls of the same 
filament-forming X-ray cavity.
Cosmic ray electrons freely diffusing isotropically from the 
center of the southern lobe of radius $r_{lobe} \approx 23$ kpc 
suggests a diffusion coefficient 
$\kappa \approx r_{lobe}^2/t_{sync} \approx 10^{30}$ cm$^2$ s$^{-1}$.
But a potential difficulty with a simultaneous origin for 
the thermal X-ray filament and southern Virgo lobe is the absence 
of a corresponding radial thermal filament in the northern lobe.

In addition, 
we now suspect that the Virgo radio lobes have a more complex evolution
resulting from a multitude of 
active galactic nucleus (AGN) outbursts
that continuously resupply cosmic ray electrons to the lobes, 
a point of view also suggested by Owen, Eilek \& Kassim. 
The 8-shaped Virgo lobes may represent an approximately bipolar 
energy ejection pattern from the central black hole in M87. 
AGN jet and cavity activity in Virgo 
(as in Perseus and many other clusters) rarely occurs along a fixed
bipolar axis as defined for example by an unchanging black hole spin,
but jet energy may be distributed in directions near such an axis. 
As a consequence, the Virgo lobes may expand much slower 
and be older than 
implied by free cosmic ray diffusion. 

With this in mind, we draw attention in Figure 1 to the sharp 
transition at the outer boundary of the Virgo radio lobes where 
the radio flux abruptly decreases, 
completely unlike the smooth gaussian transition expected 
for free diffusion with $\kappa \sim 10^{30}$ cm$^2$ s$^{-1}$. 
Owen, Eilek and Kassim (2000) emphasize that the sudden 
decrease in radio flux at the lobe boundary is real, 
not an artifact of image-processing software.
Another curious widespread feature just at this 
lobe-cluster gas transition are the bright
synchrotron-emitting rims visible in Figure 1 
immediately adjacent to the sudden sharp drop to zero radio flux.
According to Owen, Eilek and Kassim, 
this counterintuitive radio limb-brightening 
is present in over half of the perimeter of both lobes.

In the following discussion we propose that the bright 
radio rims can be understood if the diffusion coefficient of 
synchrotron-emitting electrons 
decreases sharply near the lobe-cluster gas boundaries.

\section{Plane Parallel Steady Diffusion Front}

We assume that cosmic rays (relativistic electrons and/or protons) 
in the radio lobes 
are coupled to the gas by means of a magnetic field
frozen into the gas through which cosmic rays may diffuse.
While the magnetic field allows cosmic ray pressure gradients
to act on the cluster gas, the small magnetic energy density is
dynamically negligible.
Magnetic fields of strength 0.3-10 $\mu$G are ubiquitous
in cluster gas (Govoni \& Feretti 2004). 
In general the magnetic field decreases with cluster radius
and decreasing gas density.
An approximate powerlaw relation $B \propto n_e^s$ with
$0.4 \lta s \lta 0.7$
in the Coma cluster gas 
has been established from Faraday depolarization
observations of cluster galaxies (Bonafede et al. 2010).
However, fields with energy densities
$B^2/8\pi = 6\times 10^{-13} (B/4\mu {\rm G})^2$ erg cm$^{-3}$
cannot significantly influence the dynamics of local cluster gas with
a much larger thermal energy density
$3P/2 = 5 \times 10^{-11}(n_e/0.01~{\rm cm}^{-2})                                   
(T/{\rm keV})$ erg cm$^{-3}$, and
$B^2/8\pi << 3P/2$ probably holds at every radius in the cluster gas.
In view of this, magnetic fields need not explicitly appear 
in the dynamical equations:
\begin{equation}
{ \partial \rho \over \partial t}
+ {\bf \nabla}\cdot\rho{\bf u} = 0
\end{equation}
\begin{equation}
\rho \left( { \partial {\bf u} \over \partial t}
+ ({\bf u \cdot \nabla}){\bf u}\right) =
- {\bf \nabla}(P + P_c)
\end{equation}
\begin{equation}
{\partial e \over \partial t}
+ {\bf \nabla \cdot u}e = - P({\bf \nabla\cdot u})
\end{equation}
\begin{equation}
{\partial e_c \over \partial t}
+ {\bf \nabla \cdot u}e_c = - P_c({\bf \nabla\cdot u})
+ {\bf \nabla\cdot}(\kappa{\bf \nabla}e_c)
\end{equation}
where $e = P/(\gamma - 1) = 3P/2$ and $e_c = P_c/(\gamma_c - 1) = 3P_c$ 
are the energy density-pressure relations for gas and cosmic rays
respectively 
and $\kappa$ is the difussivity of cosmic rays. 
Weak magnetic fields evolve passively, advecting with the gas. 
These equations are discussed further in Mathews \& Brighenti
(2008a,b).

Recognizing the relatively short time for gas to flow 
past the radio lobe boundary, we do not consider the energy 
lost by thermal or non-thermal radiation 
as gas flows through the boundary rims.
The time required for gas flowing 
at velocity $u \approx 20$ km s$^{-1}$ to cross the bright 
ratio rims of thickness $\Delta x \la 0.6$ kpc is 
$t_{flow} = \Delta x/u \la 3\times 10^7$ yrs.
However, the radiative cooling time in the ambient gas 
near the radio lobes 
($\rho \approx 2.5\times 10^{-26}$ g cm$^{-3}$, 
$T \approx 2.5\times 10^7$K)
is much longer, 
$t_{cool} \approx 5 m_p kT/\Lambda(T) \mu \rho 
\approx 2.7\times 10^9$ yrs, where $\Lambda(T)$ is the usual 
coefficient for optically thin thermal emission. 
Furthermore, 
the calculation we consider is embedded in 
the gravitationally
supported hot gas atmosphere in a galaxy cluster
where local cooling after the loss of energy 
by radiative emission does not occur.
Any decrease in the local gas temperature due to radiative losses  
is rapidly reversed by a global compression 
and gas inflow that maintains the virial temperature and pressure 
required to support the entire atmosphere. 
Moreover, no thermal jump associated with the Virgo radio lobe 
boundaries has been observed 
in detailed X-ray observations (Million et al. 2010).
Non-thermal electrons of energy 
$\gamma = E/m_ec^2 
\approx (2\pi m_e c^2/3B\lambda)^{1/2} \approx 6\times 10^3$ 
radiating in a $3\mu$G field 
produce synchrotron emission at $\lambda = 90$ cm as in Figure 1.
The synchrotron lifetime for these electrons,  
$t_{sync} = \gamma/{\dot \gamma} 
= (3m_e c /4 \sigma_T)(4\pi/B^2 \gamma) \approx 1.1\times 10^8$ 
yrs,
also exceeds $t_{flow}$.
However, the local synchrotron lifetime is not particularly 
relevant for our diffusion front calculation where 
new cosmic ray electrons are continuously provided 
by diffusion from the adjacent radio lobe. 
But we do assume that the synchrotron radio spectral index $\alpha$,
appropriate for a power law distribution of electron energies, 
remains constant across the diffusion front. 
Finally, since the flow time across the diffusion front at the 
lobe boundary is 
almost certainly less than the dynamical age of the lobe itself,
the diffusion front structure is approximately in steady state 
during times $\sim t_{flow}$.

For simplicity, and consistency with current Virgo X-ray observations, 
we consider the cluster gas to be isothermal 
in the vicinity of the radio lobe-cluster gas boundary, 
with pressure, $P = \rho c_s^2$, 
depending only on the isothermal sound speed 
which replaces equation (3).
The cosmic ray energy density $e_c$ is integrated 
over the relativistic energy or momentum distribution,
$e_c \propto \int EN(E) dE \propto \int p^4 f(p)(1+p^2)^{-1/2} dp$ 
and may refer to electrons and/or protons.

For steady-state, one-dimensional, plane-parallel, isothermal flow 
near the lobe-gas boundary these equations reduce to:
\begin{equation}
{d (\rho u) \over d x} = 0
\end{equation}
\begin{equation}
{d \over d x}[\rho u^2 + P + P_c] = 0
\end{equation}
\begin{equation}
{d (e_c u) \over d x}
+ P_c {du \over dx}
- {d~ \over dx}\left( {\kappa} {d e_c \over dx}\right) = 0.
\end{equation}
Equations (5) and (6) are directly integrable
\begin{equation}
\rho u = A~~~{\rm and}~~~[\rho(u^2 + c_s^2) + P_c] = B
\end{equation}
where $u<0$ and $A < 0$.
For a steady diffusion front propagating in the positive $x$ direction,
in our calculation, performed in the rest frame of the front, 
the gas flows in the negative $x$ direction 
from the cluster gas toward the lobe 
-- the direction of increasing $x$ is shown with the small
arrow in Figure 1.
Eliminating $\rho$ between equations (8) 
results in a quadratic for $u$
\begin{equation}
u^2 + u {1\over A}\left( {e_c\over3} -B\right) + c_s^2 = 0
\end{equation}
for which the desired solution is
\begin{equation}
u = {1\over2}\left\{ - {1 \over A}\left( {e_c \over 3} - B\right)
+ \left[ {1 \over A^2}\left( {e_c \over 3} - B\right)^2 
- 4c_s^2\right]^{1/2}\right\}.
\end{equation}

By differentiating equation (9)
\begin{equation}
{du \over dx} = - {u \over 3b}{de_c \over dx}~~~{\rm where}~~~
b = 2Au + [(e_c/3) - B],
\end{equation}
equation (7) can be written as
\begin{equation}
u{de_c \over dx} - {4 e_c u \over9b}{de_c \over dx}
-{d~ \over dx}\left( {\kappa} {d e_c \over dx}\right) = 0.
\end{equation}
The $x$-dependence of $\kappa$ can be subsumed into a new
independent variable 
\begin{equation}
\xi = \int {dx  \over \kappa}
\end{equation}
by multiplying each term of equation (12) by $\kappa$,
\begin{equation}
u{de_c \over d\xi} - {4 e_c u \over9b}{de_c \over d\xi}
-{d^2e_c \over d\xi^2} = 0
\end{equation}
which can be solved as a pair of first order equations
for dependent variables $\phi \equiv de_c/d\xi$ and $e_c$,
\begin{equation}
{d\phi \over d\xi} = u\phi\left( 1 - {4e_c\over9b}\right)
~~~~{de_c \over d\xi} = \phi.
\end{equation}
Once $e_c(\xi)$ is known, $u(\xi)$ follows from (10), 
and the density $\rho(\xi) = A/u(\xi)$.

Solutions of these equations describe a steady
state 1D diffusion front representing the dynamical encounter 
between cluster gas containing cosmic rays inside the radio lobe
on the small-$\xi$ side 
and approaching external
cluster gas on the large-$\xi$ side containing no cosmic rays. 
The gas pressure
must increase in the $\xi$-direction 
by $P_c = e_c/3$ to keep the total pressure
$P_t \approx P + P_c$ nearly uniform 
in subsonic flow, $u << c_s$. 
Since no X-ray cavities are known to be associated with 
the Virgo radio lobes, it is likely that the cluster gas 
pressure contributes substantially within the lobes,
i.e. the Virgo lobes are partial X-ray cavities, 
only modestly depleted of cluster gas.
The plane-parallel 
equations are valid within a distance from the lobe boundary 
that is small compared to the 23 kpc radius of the Virgo lobes. 
The solution profiles in $\xi$ -- 
$e_c(\xi)$, $\rho(\xi)$, $u(\xi)$, $P(\xi)$, etc. --  
are universal, valid for any assumed spatial variation of 
the cosmic ray diffusion coefficient $\kappa(x)$.

Without loss of generality, 
we assume that $\xi = 0$ at some arbitrary position 
$x = 0$ within the Virgo radio lobe 
where gas with temperature
$T_0 = 2.5\times10^7$ K and density 
$\rho_0 = 2.5\times10^{-26}$ gm cm$^{-3}$
($n_{e,0} = 0.0129$ cm$^{-3}$)
flows away from the front with velocity $u_0 < 0$ 
and with a cosmic ray to gas pressure ratio $(P_c/P)_0$ from
which $e_{c,0}$ can be found. 
Our integration of equations (15) begins 
inside the radio lobe at $\xi = 0$, 
where flow variables have subscript ``0'', 
and proceeds upstream toward the diffusion front at larger $\xi$.
The initial gas pressure 
$P_0 = \rho_0c_s^2$. cannot be changed much since the density
and temperature (sound speed $c_s = 580$ km s$^{-1}$)
are chosen to approximately match those 
observed in the Virgo cluster near the lobes 
(Ghizzardi, et al. 2004).
The velocity $u_0$, cosmic ray energy density $e_{c,0}$ and its
slope $\phi_0 = (de_c/d\xi)_0$ at $\xi = 0$
can be changed to alter the morphology of the front.
The numerical integration of equations (15) 
proceeds from $\xi = 0$ toward 
positive $\xi$ until $e_c$ becomes very small.

We consider two representative 
subsonic flow velocities $u_0$ at $\xi = 0$.
The first is a characteristic velocity for old Virgo lobes, 
$u_0 \approx -23{\rm ~kpc}/10^9{\rm yr} \approx -20$ km s$^{-1}$ 
and the second is somewhat larger,
$u_0 = -100$ km s$^{-1}$.
The upper three panels of 
Figure 2 show these two solutions as functions of the 
universal $\kappa$-independent variable $\xi$.
Notice that as $e_c$ approaches zero at some finite $\xi = \xi_1$ 
at the leading edge of the diffusion front, 
the gas velocity is asymptotically finite 
and negative as cluster gas enters the front; 
the second term in equation (14) also vanishes at $\xi_1$,
indicating that diffusion at the leading edge of the 
front is balanced by advection downstream.
We find $\xi_1 = 2.65\times 10^{-8}$ and $3.13\times 10^{-8}$ s
cm$^{-1}$ for $u_0 = -20$ and $-100$ km s$^{-1}$ 
respectively.

For the limiting case of a static lobe-cluster gas front where
$u \rightarrow 0$, the total pressure $P_t = P + P_c$ is
everywhere exactly uniform and equation (14) becomes simply
\begin{equation}
{d \over d\xi} \left({d e_c \over d\xi}\right) = 0.
\end{equation}
The solution in this case is linear 
$e_c(\xi) = e_{c,0} + \phi_0\xi$ with uniform slope
$de_c/d\xi =\phi_0 < 0$.
The gas pressure $P(\xi) = P_t - e_c/3$ and density 
$\rho(\xi) = [P_t - (e_c/3)]/c_s^2$ profiles are also linear.
Since the flow velocities we consider $|u_0|$ are much 
smaller than the sound speed $c_s$, the solutions plotted
in Figure 2 resemble those of static fronts.
These idealized static solutions are not
solutions of the time-dependent diffusion equation 
and represent a mathematical limit 
that may not be physically relevant.

Steady state cosmic ray diffusion fronts occur only in a 
gasdynamical environment. 
In free isotropic cosmic ray diffusion in a stationary gas
({\bf u} = 0) solutions of the diffusion equation
$\partial e_c / \partial t = {\bf \nabla\cdot}(\kappa{\bf \nabla}e_c)$
depend only on a similarity variable $\zeta = r^2/t$
and have no steady state solution.
However in Figure 2 we see that steady 
solutions are indeed possible in a fluid environment.

\subsection{Radio Synchrotron Emissivity}

Computation of the 
synchrotron emission at a fixed radio frequency 
in the diffusion front requires 
knowledge of the density of relativistic electrons and the 
magnetic field strength and orientation.
Radio polarization observations of Virgo 
at 2.8 cm by Rottmann et al. (1996) 
shown in Figure 3 reveal a remarkably high degree of 
linear polarization ($\ga 70$\%), particularly near 
the outer edges of the lobes 
where the field direction is highly ordered. 
Field vectors plotted in Figure 3 indicate that the direction of 
the magnetic field is tangent to the 
boundary of both lobes over most of the lobe perimeter. 
This is consistent with the relative compression   
in Figure 2 near the lobe-cluster gas boundary at $\xi = \xi_1$.
In our calculation the magnetic field evolves passively with 
the gas but does not have sufficient energy density to 
influence the flow parameters. 
In a one-dimensional flow with approximately 
perpendicular field alignment, magnetic flux
is conserved if $B \propto \rho$.

For a power law distribution of relativistic electron energies,
$N(E)dE = KE^{-p}dE$ with $p > 2$ the total energy density of
cosmic ray electrons is
\begin{equation}
e_c = \int_{E_{min}}^{E_{max}} EN(E)dE = {K\over (p-2)}
(E_{min}^{2-p}-E_{max}^{2-p})
\end{equation}
\begin{displaymath}
~~~~~~~~~~~~~~~~~\approx {K\over (p-2)}E_{min}^{2-p}
\end{displaymath}
and therefore $K \propto e_c E_{min}^{p-2}$.
As the magnetic field slowly varies across the diffusion front,
the relativistic adiabatic invariance of the 
electron magnetic moment
requires that $\beta_{\perp}\gamma \propto B^{1/2}$
(Sturrock 1994) 
so $E_{min} \propto B^{1/2}$
and $K \propto e_c B^{{1\over2}(p-2)}$.
For electrons with intermediate energies,
$E_{min} << E << E_{max}$,
the synchrotron emissivity from the electron distribution is
\begin{equation}
j_{\nu} \propto KB (\nu/B)^{-\alpha}
\end{equation}
(Longair, 1994) 
where we assume $p = 2.5$ and $\alpha = (p-1)/2 = 0.75$.
Combining these dependencies, 
the radio synchrotron emissivity observed
at the same frequency over the whole profile
should vary with $e_c$ and $\rho$ as
\begin{equation}
j_{\nu} \propto e_c B^{{1\over2}(p-2)} \cdot
B \cdot B^{{1\over2}(p-1)} \propto e_c \rho^{\beta}
\end{equation}
where
\begin{equation}
\beta = {1 \over 2}(2p - 1) = {1 \over 2}(4\alpha + 1)
\end{equation}
so $\beta = 2$ when $p = 2.5$. 
Owen, Eilek \& Kassim (2000) suggest a mean index 
$\alpha \approx 1.0$ for Virgo for which $\beta = 2.5$.
The index $\alpha$ may be even steeper near the lobe boundaries 
or vary with $x$, but current observations do not warrant 
such refinements.

What is the condition that the emissivity 
$j_{\nu} \propto e_c \rho^{\beta}$ (where $\beta \approx 2$) has a
limb-brightened maximum at some $\xi$ in the diffusion front?
Since $e_c$ always tends to zero as $\xi \rightarrow \xi_1$ 
near the leading edge
of the diffusion front, a maximum is guaranteed if
the emissivity increases with $\xi$ at $\xi = 0$, i.e.
\begin{equation}
{1 \over j_{\nu}}{dj_{\nu} \over d\xi} =
{1\over e_c} {d e_c \over d\xi}
+ \beta {1\over \rho}{d\rho \over d\xi} > 0
\end{equation}
at $\xi = 0$.
Using $A = \rho u$ and equation (11), 
the density derivative in the second term can be written 
\begin{equation}
{1\over \rho}{d\rho \over d\xi} =
-{1\over u}{du \over d\xi} =
{1 \over 3b}{de_c \over d\xi}.
\end{equation}
The condition for a maximum in $j_{\nu}$ is then
\begin{equation}
{\phi \over e_c} + {\beta \phi \over 3b} =
{\phi \over e_c}
\left[1 + \beta {P_c \over \rho u^2 -\rho c_s^2}\right] > 0
\end{equation}
at $\xi = 0$.
Since $\phi_0 = (de_c/d\xi)_0$ is always negative, the
quantity in brackets must also be negative.
Furthermore, we expect diffusion fronts to be highly subsonic,
$u^2 << c_s^2$, so the condition for a synchrotron emission
maximum in the diffusion front is
\begin{equation}
{P_c \over P} > {1 \over \beta} = {2 \over 4\alpha + 1} 
~~~{\rm or}~~~
{P_c \over P +P_c} > {1 \over 1+\beta} = {2 \over 4\alpha + 3} 
\end{equation}
inside the lobe at $\xi = 0$.
Normalized emissivity profiles $j_{\nu}(\xi)/j_{\nu}(0)$ 
with $\beta = 2$
are plotted in the lowest panel of Figure 2 
for the two chosen values of $u_0$ and 
for $(P_c/P)_0 = 1$ which produces 
the desired maximum for radio limb brightening.  

The spatial variation of $e_c$ and other
parameters with the true spatial coordinate $x$ can be
found for any assumed $\kappa(x)$ by integrating
$dx/d\xi = \kappa$ to determine $x = x(\xi)$.
If $\kappa$ is constant, then $\xi$ in
the solutions in Figure 2 can simply be replaced with $x/\kappa$.

\subsection{Diffusion Fronts in Physical Space}

From any solution in variable $\xi$ 
like those in Figure 2 
an infinite variety of solutions in 
physical space $x(\xi)$ can
be found by choosing various functions $\kappa(x) = dx/d\xi$.
In particular, very sharply decreasing profiles
$e_c(x)$ and $j_{\nu}(x)$
can be generated even when $e_c(\xi)$ and $j_{\nu}(\xi)$ 
are very slowly varying.
By this means, $x(\xi)$ could be chosen to match
the limb-brightened $j_{\nu}(x)$ and radio flux 
observed at the edge of the Virgo lobes.

Since the diffusion coefficient $\kappa$ must be positive, 
the spatial coordinate $x$
must be an increasing function of $\xi$
with positive slope $\kappa = dx/d\xi > 0$.
We consider two simple examples for $x = x(\xi)$ 
normalized so that $x(\xi = 0) = 0$: 
\begin{equation}
{\rm Case~I:}~~~~{x \over x_0} = (1 - e^{-\xi/\xi_0})
\end{equation}
\begin{equation}
\kappa = {dx \over d\xi} = {x_0 \over \xi_0}e^{-\xi/\xi_0} 
= {x_0 \over \xi_0}\left[1 - {x \over x_0}\right]
\end{equation}
and 
\begin{equation}
{\rm Case~II:}~~~~{x \over x_0} = \tanh\left({\xi \over \xi_0}\right)
\end{equation}
\begin{equation}
\kappa = {dx \over d\xi} = {x_0 \over \xi_0}
\cosh^{-2}\left({\xi \over \xi_0}\right)
= {x_0 \over \xi_0}\left[ 1 - \left({x \over x_0}\right)^2\right].
\end{equation}
For both cases the physical scale of the front 
extends from $x = 0$ to $x = x_0$ where $\kappa$ becomes zero.
The parameter $\xi_0$ characterizes the rate of distortion of 
the $\xi$-profiles into $x$-profiles with smaller values of $\xi_0$ 
concentrating the $x$-profiles closer to $x = x_0$

Figure 4 shows the variation of 
$\kappa(x)$ and $j_{\nu}(x)/j_{\nu}(0)$ for both Case I and II 
with $x_0 = 1$ kpc for three values of $\xi_0$, 
all for the $u_0 = -20$ km s$^{-1}$ solution plotted in Figure 2.
Figure 5 shows the spatial variation of the flow variables 
on the radio lobe side of the diffusion front,
$\rho(x)$, $u(x)$, $B(x)$, $P_c(x)$ and $P(x)$ 
for $\xi_0 = 3 \times 10^{-9}$ s cm$^{-1}$ and  
$x_0 = 3.08 \times 10^{21}$ cm = 1 kpc. 
The top panel shows a rapid decrease in 
$\rho(x)/\rho_0$ and $B(x)/B_0$ behind 
the front near $x = x_0$ as gas accelerates away
from the front into the lobe.
The decreasing field strength in this accelerating 
flow is responsible for the narrowness of the 
observed limb brightened region at the lobe boundary.
As $x$ approaches $x_0$ from the lobe side, 
the cosmic ray diffusion coefficient $\kappa(x)$
(really $\kappa_{\perp}(x)$) plummets
(along with $P_c(x)$) toward zero.
On the cluster gas side of the front, $x > x_0$,
where $e_c$ is assumed to vanish, 
equation (9) requires that 
the flow velocity $u$ remains locally uniform, 
$u(x > x_0) = u(x_0)$ 
along with the gas density and pressure.
In addition, the magnetic field perpendicular to the flow 
continues to be uniform in the cluster gas beyond the radio lobe;
bright radio rims occur as the diffusing electrons 
encounter this inflowing field.
Evidently the tangential $B$ field just ahead of 
the front is prepared in a compression 
between a subsonically outwardly expanding lobe
and/or radiatively cooling cluster gas inflowing toward the lobe.


\subsection{X-ray Features Associated with Radio Lobes}

The pressure of relativistic electrons 
responsible for the radio emission observed in Virgo 
is unlikely to be 
the only or even the dominant non-thermal pressure component.
The electron energy distribution $EN(E)$  
peaks toward lower energies that emit at radio wavelengths too 
large to be currently detected.
In studies of radio synchrotron and inverse Compton X-ray emission 
from electrons of the same energy inside X-ray cavities, 
assumed to be in pressure balance with local cluster gas,
Croston et al. (2005) find no evidence for an energetically 
dominant relativistic proton population. 
Although relativistic protons
dominate the cosmic ray energy density in the Milky Way, 
they are difficult to detect in extragalactic sources.

Observational efforts should be made to detect the 
total non-thermal pressure in the Virgo radio lobes 
by identifying a lower gas pressure in the lobe 
compared to that in cluster gas outside the lobes 
at the same distance from the cluster center.
Small irregularities in the azimuthally-averaged 
radial X-ray surface brightness profile in Virgo have 
been discussed in detail (Churazov et al. 2008),
but a similar effort should be made to detect azimuthal  
variations in the X-ray surface brightness and gas pressure 
that correlate with the radio lobe structures.
Since the radio lobe boundaries are well-defined, 
it should be possible to detect or set rather tight limits on 
the contribution of non-thermal components to the 
pressure in the lobe.

To explore this possibility, 
we constructed an approximate azimuthally-averaged  
radial (bolometric) X-ray surface brightness profile 
in Virgo (using the Ghizzardi et al. 2004 observations) 
and estimated the brightness reduction produced by 
depleting the gas density by two 
inside symmetric radio lobes of radius 23 kpc 
with centers 13.7 kpc from the cluster center in M87, 
assumed to be in the plane of the sky.
We found that the X-ray surface brightness in mid-lobe 
regions is depleted by about 0.55 - 0.70 along M87-centered
circles of radius 25 - 30 kpc, 
circles that also contain large regions 
of cluster gas outside the lobes that can be used for calibration. 
Near the lobe rims the X-ray surface brightness profile
is likely to have a slope change, not a sharp discontinuity. 
The gas temperature may also differ across the lobe boundary
since thermal conductivity will also be reduced
by magnetic fields tangent to the boundary.

However, no lowering of the gas pressure 
in the radio lobes is 
apparent in the recent Virgo pressure 
image of Million et al. (2010), although low amplitude thermal 
pressure variations may eventually be detected with 
dedicated X-ray observations. 
If the contribution of cosmic rays to the total pressure 
is lower than we assume here, $P_c/(P+P_c) \approx 1/3$,
the condition (24) for radio limb brightening requires 
a larger value of the spectral index $\alpha$. 
For example, if $P_c$ contributes only about 13\% of the 
total pressure in the lobes, 
we require $\alpha \ga 3$ for limb-brightening, 
but the structure of the diffusion 
front remains similar to those in Figures 2, 4 and 5. 

The terminology for feedback-related 
diffuse radio emitting regions in 
galaxy clusters includes mini-halos, lobes and emission 
from X-ray cavities.
We stress here that all of these extended  
radio-emitting regions must to some extent also be 
X-ray cavities, although may not have been 
detected with current X-ray observations 
particularly if $P_c \ll P$.
Another complication that can arise 
in detecting X-ray cavities is the possibility of 
inverse Compton X-ray emission from upscattered cosmic 
microwave radiation, as recently observed in the large 
radio lobes in the Fornax cluster
(Tahiro et al. 2009; Seta et al. 2011).
Additional inverse Compton X-radiation that accompanies 
the radio emission would make detection of thermal 
cavities in the hot cluster gas more difficult. 

\section{Final Remarks}

Our simple calculation shows that
cosmic rays in the Virgo cluster can be strongly confined 
within certain regions of the cluster gas. 
Confinement is aided by magnetic fields that are  
tangent to the boundaries of these regions 
where the synchrotron emission decreases with 
decreasing cosmic ray diffusivity.
Broad gaussian-shaped diffusion fronts 
associated with free diffusion 
at constant diffusivity $\kappa$ 
in the cluster gas are not present in Virgo 
and cannot be generally assumed. 
Nevertheless, free diffusion of synchrotron emitting electrons 
may occur in nature and can be detected by 
spectral {\it flattening} near the outer edges of 
gaussian-broadened radio lobes 
because the diffusion coefficient increases for 
more energetic relativistic electrons.

The radio lobes in Virgo, and those in other clusters, 
are naturally buoyant because the additional 
partial pressure of cosmic rays depresses the thermal gas pressure 
(and gas density) in the lobe to match the cluster gas pressure 
just outside the lobe.
As these large buoyant lobes attempt to slowly rise in the 
cluster atmosphere, they encounter cluster gas that is 
slowly inflowing due to radiation losses, 
compressing gas throughout the lobes and 
at the lobe-cluster gas boundaries.
This compression forces the cluster magnetic field to 
become nearly perpendicular to the direction of compression, 
and tangent to the diffusion front interface, 
sharply limiting cosmic ray diffusion in this direction. 
In this sense 
the propagation of cosmic rays in cluster gas is self-limiting. 

Except for the idealized singular case where the magnetic field 
is aligned precisely perpendicular to the diffusion front,
as the expanding radio lobe and inflowing cluster 
gas mutually compress at glacial velocities, 
the magnetic field is forced to become 
tangent to the lobe boundary, becoming locally perpendicular 
to the compression regardless of its initial morphology.
Evidently, a similar field-orienting compression 
occurs throughout the Virgo radio lobes, 
as supported by the polarization vectors in Figure 3. 
Such an alignment cannot simply be due to 
so-called ``draping'' of magnetic lines of force 
along its boundary since it is 
most unlikely that large scale magnetic field lines 
exist that connect across the lobe boundaries
and throughout the interior of the lobes.
Considering the dynamic activity inside the lobes 
visible in Figure 1,
the spatial continuity of polarization vectors in Figure 3 
from one observed beam position to the next 
along the field direction almost certainly  
does not trace a single magnetic line of force. 
Instead, we envision a field that naturally 
aligns perpendicular to a large scale one-dimensional compression 
much like spaghetti arranges itself horizontally 
when served on a dinner plate.
Magnetic fields {\it of all spatial scales} are aligned in this manner, 
accounting for the remarkable consistency 
of the direction 
and magnitude of the polarization vectors observed in Virgo.
The insensitivity of compressive field alignment 
to the spatial scale or local curvature of the field 
is particularly relevant 
to galaxy clusters like Virgo 
where the relativistic electron gyroradius 
$r_g = \gamma mc^2/eB = 5\times 10^{12}(\gamma/10^4)
(B/3\mu{\rm G})^{-1}$ cm 
is $10^{10}$ times smaller than the radius of the lobes.
The polarizing magnetic field
experienced by synchrotron-radiating
electrons, when averaged over kpc-sized radio beams as in 
Figure 3, gives the impression of magnetic field lines  
that are coherent over kpc-scales, 
i.e. that single field lines are ``draped''.
However, the field seen by the radiating electrons 
is likely to have 
a range of different scales (radius of curvature) but the electrons
responsible for the observed polarized emission radiate from field
segments that are locally aligned perpendicular to both the line of
sight and to the large scale compression The radiating segments do not
need to be (and are probably not) connected along a single large scale
line of force.

The region of compressionally aligned $B$ in our diffusion fronts 
could extend significantly beyond 
the lobe boundary where $j_{\nu} \rightarrow 0$ in Figure 1. 
A similar cosmic ray confinement by boundary-tangent ${\bf B}$
is also expected in X-ray cavities 
with corresponding radio continuum 
limb-brightening and high polarization 
near the radio emission boundaries at the cavity walls.
Detailed radio observations of magnetic field polarization 
and alignment along the walls of known X-ray cavities 
do not appear to have been attempted.

Our steady state solutions describe the upstream diffusion of  
relativistic cosmic rays from a radio lobe slowly compressing 
against cluster gas where the cosmic ray energy density is negligible. 
As indicated by radio polarization observations of Virgo, 
the magnetic field in the upstream cluster gas 
is aligned largely 
tangent to the lobe boundary and perpendicular to the gas flow.
At the leading edge of the advancing cosmic ray diffusion front, 
where non-thermal particles first penetrate
into the undisturbed cluster gas, 
the diffusion coefficient declines sharply 
with increasing magnetic field strength 
and its unfavorable alignment. 
Nevertheless, the small non-thermal pressure introduced by diffusing 
cosmic rays near the leading edge of the front lowers the 
local gas pressure and the reduced 
gas density $\rho = P/c_s^2$ drives a gas flow that expands 
back toward the radio lobe in a direction opposite to the 
forward advancing diffusion. 
Under certain conditions bright radio synchrotron rims are expected 
along the diffusion front at the lobe boundary 
as shown in Figure 1.
As the steady diffusion front is approached from the radio lobe side,
the synchrotron emissivity rises because of the increasing 
magnetic field ($B \propto \rho$) but ultimately 
declines with the decreasing number density of 
synchrotron-emitting electrons.
The sharpness of the radio synchrotron emission peaks 
and its decline to zero flux  
depends on the rate that the diffusion coefficient 
decreases across the diffusion front.
Bright radio synchrotron rims are expected when the partial 
pressure of cosmic rays inside the radio lobe 
and the spectral index are sufficiently large. 

Our analysis in this paper describes lobe rim profiles 
that are in qualitative agreement with the radio image of Virgo.
As more data become available at several radio frequencies,
a precise fit to the radio limb brightening at the 
lobe boundaries in Virgo 
may allow for the first time an estimate of the spatial profiles 
of the field strength $B(x)$ and the corresponding 
cosmic ray diffusion coefficient $\kappa(x)$. 
In addition, it may be possible to determine 
from X-ray observations whether the total non-thermal pressure 
in the Virgo lobes can be estimated from azimuthal gas density
and pressure variations across the lobes along 
cluster-centered circular arcs that intersect both the radio 
lobes and cluster gas outside the lobes.
Indirect detection or non-detection 
of the pressure of additional unobserved thermal or 
relativistic particles in radio lobes would have 
broad astrophysical implications. 
Finally, we note that the presence of a gas density jump at
the lobe-cluster gas transition in Virgo 
as in Figure 5 may help to divert
and divide the outflowing gas at the top of the eastern 
mushroom-like flow visible in Figure 1.


\vskip.1in
\acknowledgements
We thank Fill Humphrey for discussions about azimuthal 
variations in the X-ray emission from Virgo. 
Studies of the evolution of hot gas in elliptical galaxies
at UC Santa Cruz are supported by NSF and NASA grants 
for which we are very grateful.

                                                                                   
\clearpage

\begin{figure}
\vskip1.in
\centering
\includegraphics[width=4.in,scale=0.8,angle=0]{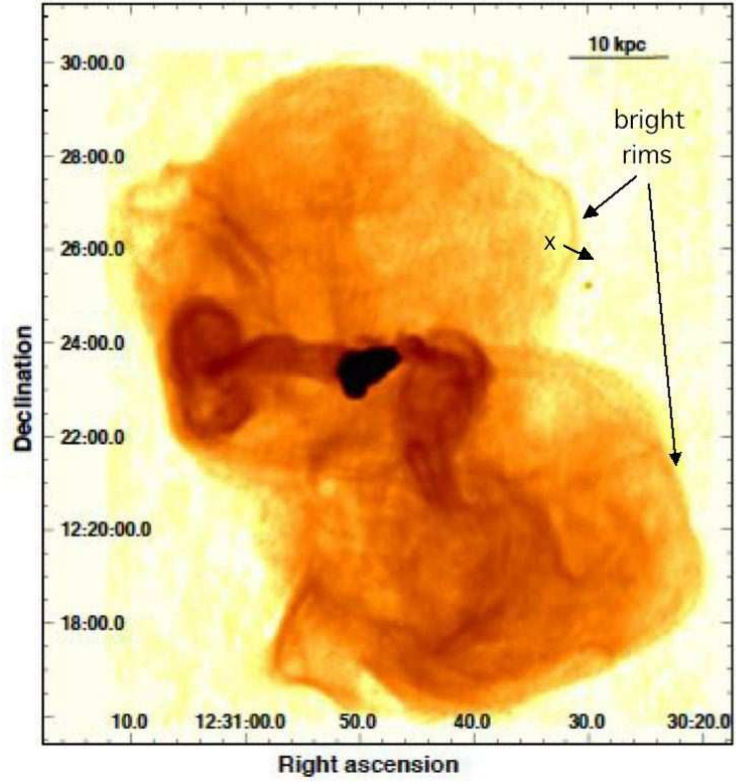}
\caption{
VLA radio image of Virgo at 90 cm (Owen, Eilek, Kassim 2000)
showing radio continuum limb-brightening 
(beam resolution is $0.6$ kpc). The small arrow 
at the western rim of the northern lobe shows the direction 
of increasing $x$ in the diffusion front.}
\label{f1}
\end{figure}

\clearpage

\begin{figure}
\vskip.5in
\centering
\includegraphics[width=3.in,scale=1.9,angle=0]{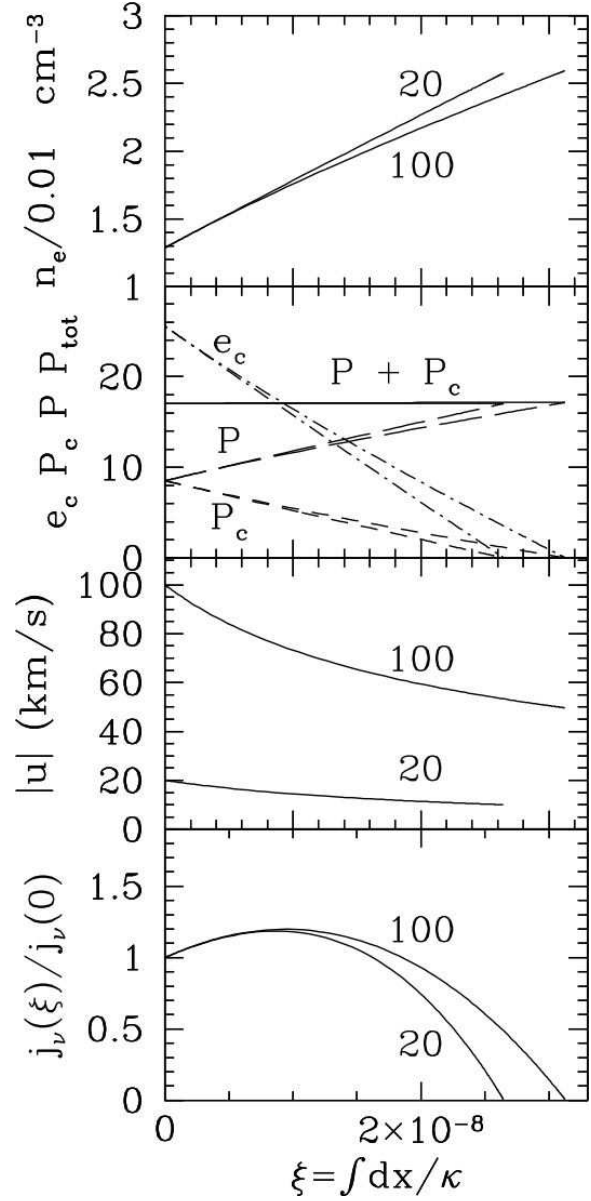}
\caption{
Two solutions of equations (15) with $u_0 = -20$ and $-100$ 
km s$^{-1}$, both with 
$T = 2.5\times10^7$ K, $\rho_0 = 2.5\times 10^{-26}$ gm cm$^{-3}$
(or $n_{e,0} =0.0129$ cm$^{-3}$), 
$(P_c/P)_0 = 1$, and $\phi_0 = -0.01$. 
The second panel from the top 
shows solutions for $e_c(\xi)$ (dashed-dot lines), 
$P_c(\xi)$ (short dashed lines), 
$P(\xi)$ (long dashed lines) and 
$P_t = P + P_c$ (solid lines). 
Solutions for $u_0 = -100$ extend to slightly larger $\xi$ 
than those for $u_0 = -20$ km s$^{-1}$.
Bottom panel is the profile $j_{\nu}(\xi)$ of the 
radio synchrotron emissivity normalized to unity at $\xi = 0$ 
and with $\beta = 2$.
}
\label{f2}
\end{figure}


\clearpage

\begin{figure}
\vskip1.in
\centering
\includegraphics[width=4.in]{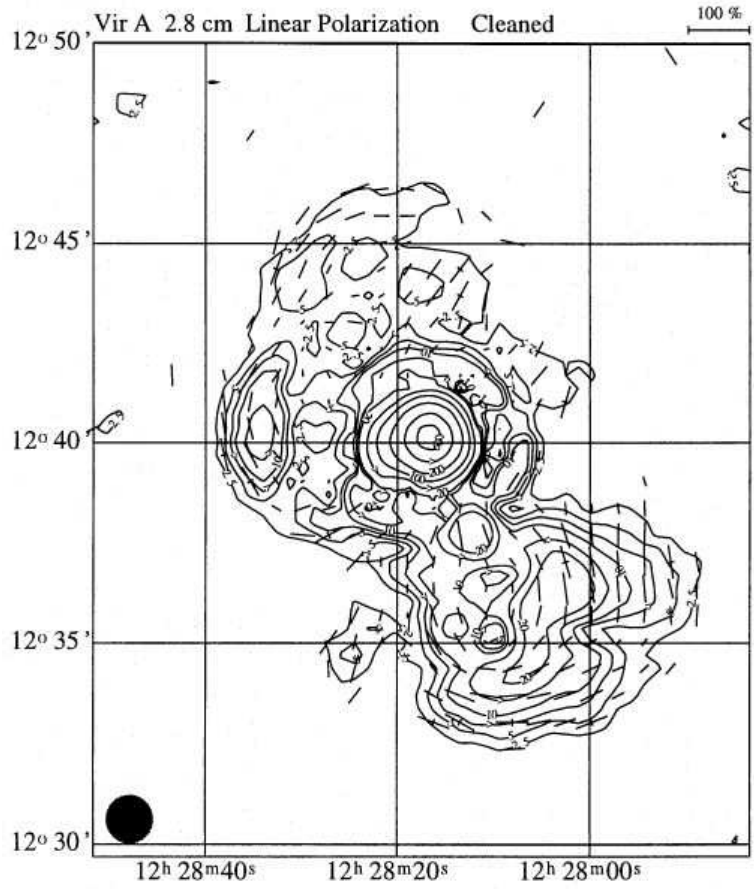}
\vskip0.2in
\caption{
Polarized intensity map of Virgo at 2.8 cm 
with superimposed B vector
with lengths proportional to the degree of linear polarization.
Circle at lower left shows beam size 
(Rottmann, et al. 1996).
}
\label{f3}
\end{figure}

\clearpage

\begin{figure}
\vskip1.in
\centering
\includegraphics[width=3.in]{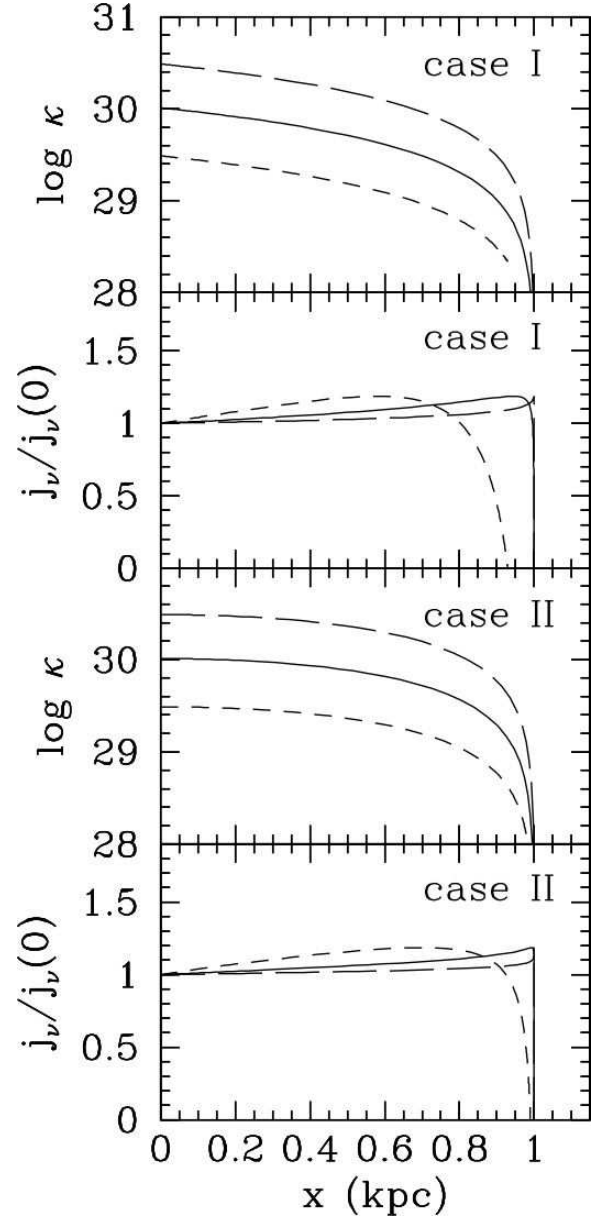}
\vskip0.2in
\caption{
Normalized radio synchrotron emissivity profiles 
$j_{\nu}(x)/j_{\nu}(0)$ for the $u_0 = -20$ km s$^{-1}$ flow
in physical space for Case I and II diffusion 
coefficients $\kappa(x)$ (cm$^2$ s$^{-1}$) 
both with $x_0 = 1$ kpc. 
The three profiles in each panel correspond to 
$\xi_0 = 1 \times 10^{-9}$ ({\it long dashed line}),
$\xi_0 = 3 \times 10^{-9}$ ({\it solid line}),
and 
$\xi_0 = 1 \times 10^{-8}$ ({\it short dashed line}), 
all in units of s cm$^{-1}$.
}
\label{f4}
\end{figure}

\clearpage

\begin{figure}
\vskip1.in
\centering
\includegraphics[width=3.in]{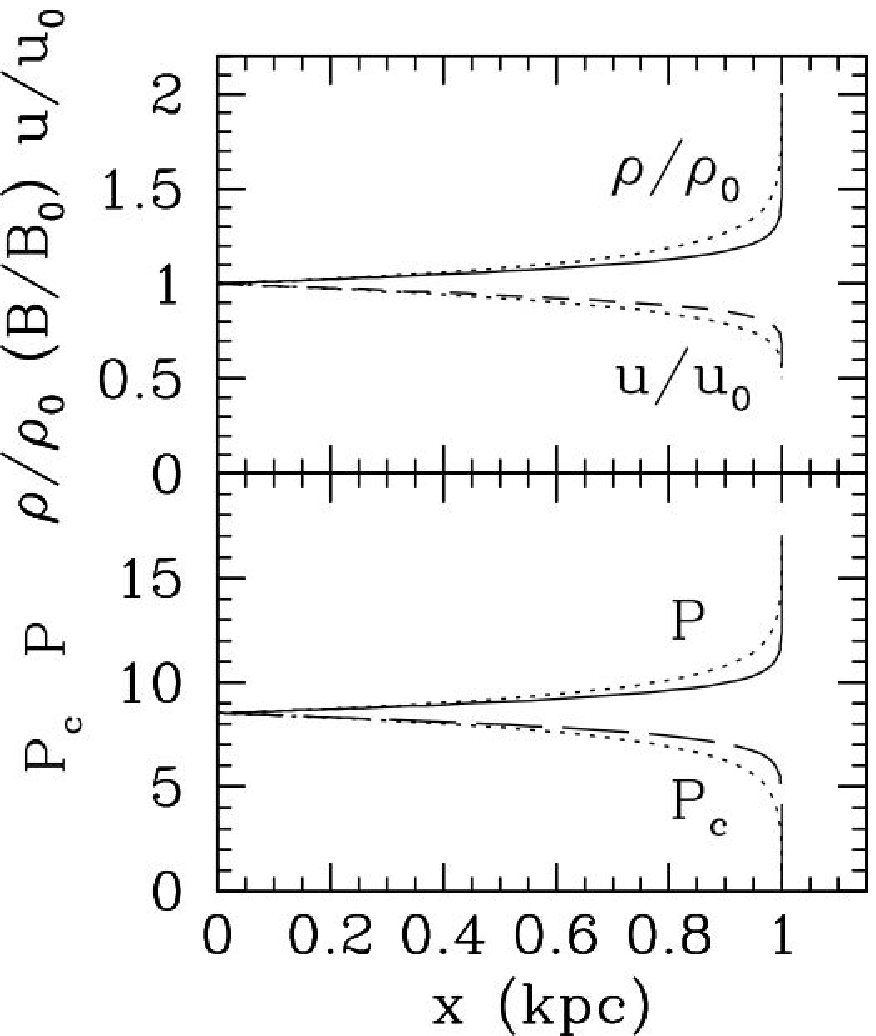}
\vskip0.2in
\caption{
Diffusion front profiles for $u_0 = -20$ km s$^{-1}$ flow 
with $\xi_0 = 3 \times 10^{-9}$ and $x_0 = 1$ kpc. 
Coordinate transformations $x(\xi)$ for Case II are shown 
in solid and long dashed curves; adjacent dotted lines show 
profiles for Case I.
{\it Upper panel:} Gas density and magnetic field profiles
(solid line) and flow velocity (short dashed line),  
all normalized to values at $x = 0$.
{\it Lower panel:} gas pressure $P(x)$ (solid line) and cosmic ray 
pressure $P_c(x)$ (long dashed line) both in units of 
$10^{-11}$ dy cm$^{-2}$.
}
\label{f5}
\end{figure}


\begin{references}

\reference{} Bonafede, A., Feretti, L., Murgia, M., Govoni, F.,
Giovannini, G., Dallacasa, D., Dolag, K., Taylor, G. B., 2010, A\&A,
513, 30

\reference{} Churazov, E., Forman, W., Vikhlinin, A.,
Tremaine, S.,Gerhard, O. \& Jones, C., 2008, MNRAS, 388, 1062

\reference{} Croston, J. H., Hardcastle, M. J., Harris, D. E.,
Belsole, E., Birkinshaw, M., Worrall, D. M., 2005, ApJ, 626, 733

\reference{} Forman, W.; Nulsen, P.; Heinz, S.; Owen, F.; Eilek, J.;
Vikhlinin, A.; Markevitch, M.; Kraft, R.; Churazov, E.; Jones, C., 
2005, ApJ, 635, 894

\reference{} Ghizzardi, S., Molendi, S., Pizzolato, F., \& De Grandi,
S. 2004, ApJ, 609, 638

\reference{} Govoni, F., Feretti, L., 2004,
International Journal of Mod. Physics D13, 1549

\reference{} Owen, Frazer N.; Eilek, Jean A.; Kassim, Namir E., 2000,
ApJ, 543, 611

\reference{} Longair, M. S., 1994, in 
``High Energy Astrophysics'', vol 2, 2nd ed., 
(Cambridge University Press: Cambridge), p. 251

\reference{} Mathews, W. G. \& Brighenti, F. 2008b, ApJ, 685, 128

\reference{} Mathews, W. G. \& Brighenti, F. 2008a, ApJ, 676, 880

\reference{} Million, E. T.,Werner, N., Simionescu, A., 
Allen, S. W., Nulsen, P. E. J., Fabian, A. C., Bohringer, H., \&
Sanders, J. S., 2010, MNRAS, 407, 2046

\reference{} Rottmann, H.; Mack, K.-H.; Klein, U.; Wielebinski, R.,
1996, A\&A, 309, L19

\reference{} Seta, H., Tashiro, M. S., Isobe, N. 2011,
Proceedings IAU Symposium No. 275,
{\it Jets on all Scales}, eds.
G. E. Romero, R. A. Sunyaev \& T. Belloni, p. 184

\reference{} Sturrock, P. A. 1994, in 
``Plasma physics: an introduction to the theory of astrophysical,
geophysical \& laboratory plasmas'',
(Cambridge University Press: Cambridge), p. 40

\reference{} Tashiro, M. S., Isobe, N., Seta, H.,
Matsuta, K., \& Yaji, Y. 2009, PASJ, 61, S327

\reference{} Werner, N. et al. 2010, MNRAS, 407, 2063

\reference{} Young, A. J.; Wilson, A. S.; Mundell, C. G., 2002,
ApJ, 579, 560



\end{references}
\end{document}